\documentclass[letter,superscriptaddress,twocolumn,prl,showkeys,showpacs]{revtex4-1}

\usepackage[utf8]{inputenc}
\usepackage{amsmath}   
\usepackage{graphicx}   
\usepackage{verbatim}  
\usepackage{color}     
\usepackage{subfigure}  
\usepackage{hyperref}   
\usepackage{natbib}
\usepackage{ gensymb }
\usepackage{color}

\begin{document}

\title{Doping nature of native defects in $1T$-TiSe$_2$}

\author{B. Hildebrand}
\altaffiliation{Corresponding author.\\ baptiste.hildebrand@unifr.ch}
\affiliation{D{\'e}partement de Physique and Fribourg Center for Nanomaterials, Universit{\'e} de Fribourg, CH-1700 Fribourg, Switzerland}

\author{C. Didiot}
\altaffiliation{Corresponding author.\\ clement.didiot@unifr.ch}
\affiliation{D{\'e}partement de Physique and Fribourg Center for Nanomaterials, Universit{\'e} de Fribourg, CH-1700 Fribourg, Switzerland}

\author{A. M. Novello}
\affiliation{D{\'e}partement de Physique de la Mati{\`e}re Condens{\'e}e, University of Geneva, 24 Quai Ernest-Ansermet, 1211 Geneva 4, Switzerland}

\author{G. Monney}
\affiliation{D{\'e}partement de Physique and Fribourg Center for Nanomaterials, Universit{\'e} de Fribourg, CH-1700 Fribourg, Switzerland}

\author{A. Scarfato}
\affiliation{D{\'e}partement de Physique de la Mati{\`e}re Condens{\'e}e, University of Geneva, 24 Quai Ernest-Ansermet, 1211 Geneva 4, Switzerland}

\author{A. Ubaldini}
\affiliation{D{\'e}partement de Physique de la Mati{\`e}re Condens{\'e}e, University of Geneva, 24 Quai Ernest-Ansermet, 1211 Geneva 4, Switzerland}

\author{H. Berger}
\affiliation{Institut de G{\'e}nie Atomique, Ecole Polytechnique F{\'e}d{\'e}rale de Lausanne, CH-1015 Lausanne, Switzerland}

\author{D. R. Bowler}
\affiliation{London Centre for Nanotechnology and Department of Physics and Astronomy,
University College London, London WC1E 6BT, UK}

\author{C. Renner}
\affiliation{D{\'e}partement de Physique de la Mati{\`e}re Condens{\'e}e, University of Geneva, 24 Quai Ernest-Ansermet, 1211 Geneva 4, Switzerland}

\author{P. Aebi}
\affiliation{D{\'e}partement de Physique and Fribourg Center for Nanomaterials, Universit{\'e} de Fribourg, CH-1700 Fribourg, Switzerland}

\begin{abstract}
The transition metal dichalcogenide $1T$-TiSe$_2$ is a quasi two-dimensional layered material with a charge density wave (CDW) transition temperature of  $T_{\text{CDW}}$ $\approx$  200 K. Self-doping effects for crystals grown at different temperatures introduce structural defects, modify the temperature dependent resistivity and strongly perturbate the CDW phase. 
Here we study the structural and doping nature of such native defects combining scanning tunneling microscopy/spectroscopy and \textit{ab initio} calculations. The dominant native single atom dopants we identify in our single crystals are intercalated Ti atoms, Se vacancies and Se substitutions by residual iodine and oxygen. 
\end{abstract}
\date{\today}
\maketitle

$1T$-TiSe$_2$ is a widely studied quasi-2D transition metal dichalcogenide. It undergoes a second-order phase transition to a charge density wave (CDW) state, taking place at $T_{\text{CDW}}\approx 200$K and leading to a commensurate superstructure\cite{salvo1976}. In addition, pressure induced superconductivity has been observed in pure $1T$-TiSe$_2$\cite{Kusmartseva2009a}.

Whereas $1T$-TiS$_2$ is a semiconductor and $1T$-TiTe$_2$ a semimetal, it is not yet clear whether $1T$-TiSe$_2$ is a semiconductor with a very small indirect gap\cite{Rasch2008} or a semimetal\cite{Greenaway1965a,Pillo2000} with a very small indirect band overlap. This uncertainty may be related to the presence of self-doping during the crystal growth. Indeed, increasing the crystal growth temperature induces Ti excess\cite{salvo1976} that strongly modifies the temperature dependent resistivity and perturbates the CDW phase. Furthermore, in the debate on the origin of the CDW a broad resistivity maximum around $T_{\text{CDW}}$ has been associated to exciton condensate formation\cite{Wilson1978a,Cercellier2007a}, and electron-hole fluctuations \cite{Monney2012b} already above  $T_{\text{CDW}}$ with strong band renormalization make the determination of the normal-state band structure difficult.

Physical properties of semiconductors (semimetals) are very sensitive to dopants and being able to control density and nature of single dopants is essential for the development of new devices\cite{Koenraad2011}. Extrinsic doping\footnote{The term of doping is taken in the general sense of the modification of electronic properties induced by the introduction (intentional or accidental) of impurities or defects into a pure compound (not necessarily in a well defined semiconductor).} of $1T$-TiSe$_2$ can lead to interesting modifications of the native electronic properties. For example, controlled intercalation of Cu atoms (4-8$\%$) induces superconductivity and reduces simultaneously the transition temperature to the CDW state, suggesting competition between these two collective electronic states\cite{Morosan2006a}. For semiconductors in general and topological insulators, where dopants can be of extrinsic or native origin\cite{Schofield2003,Kim2011,Alpichshev2012,Kendrick1996}, only few extensive investigations of structural and electronic effects of defects have been done, and such a detailed study is still missing for $1T$-TiSe$_2$\cite{Kuznetsov2012a}.

Here we focus on the identification and electronic characterization of pre-existing (native) defects in $1T$-TiSe$_2$ to provide a better understanding of the effect of Ti self-doping and extrinsic doping.
Scanning tunneling microscopy and spectroscopy (STM/STS) are perfect local tools for getting structural and spectroscopic information up to atomic resolution. A comparison with \textit{ab initio} Density Functional Theory (DFT) calculations allows to compare measurements and theoretical predictions in order to confirm them. 
We present a combined study of these techniques which enables the identification and characterization of the structural and electronic properties of native structural defects in $1T$-TiSe$_2$. This study shows the predominant native single atom dopants to be intercalated Ti atoms, Se vacancies and Se substitutions by residual iodine and oxygen atoms. Indeed, iodine is used as catalyst for the crystal growth and oxygen is the main contaminant in iodine and selenium gas.

The three $1T$-TiSe$_2$ crystals used in this study were all grown by iodine vapor transport. The first crystal was grown at 575\celsius$~$ in order to be stoichiometric (with negligible amount of Ti excess) and the second and the third one were grown respectively at 650\celsius$~$ and 900\celsius, therefore containing excess Ti up to 1.5$\%$\cite{salvo1976}. The surface was cleaved \textit{in situ} and directly measured at low temperature in ultra high vacuum (UHV) with a base pressure below 3$\cdot$10$^{-11}$ mbar. STM/S measurements were performed at 4.7 K with an Omicron LT-STM and a SPECS JT-STM. STM measurements were taken in constant current mode by applying a bias voltage to the sample. The differential conductance d$I$/d$V$ curves (STS) were recorded with an open feedback loop using the standard lock-in method (bias modulation 20 mV peak to peak at 965 Hz). 
Calculations were performed using the plane wave pseudo-potential code VASP\cite{kresse1996,kresse1993}, version 5.3.3.  Projector-augmented waves\cite{kresse1999} were used with the PBE\cite{perdew1996} exchange correlation functional and plane wave cut-offs of 211 eV ($1T$-TiSe$_2$, I substitutional) and 400 eV (O).  We used two cell sizes in our model: 12.26\AA$\times$14.16\AA (small cell) and 24.52\AA$\times$28.32\AA (large cell).  The $1T$-TiSe$_2$ surface was modelled with four layers (for small cells) or two layers (for large cells) with the bottom layer of Se fixed.  A Monkhorst-Pack mesh with $4 \times 4 \times 1$ and $2 \times 2 \times 1$ $k$-points was used to sample the Brillouin zone of the small cell and the large cell, respectively.  The parameters gave energy difference convergence better than 0.01 eV.  During structural relaxations, a tolerance of 0.03 eV/\AA\ was applied. STM images were generated using the Tersoff-Hamann\cite{Tersoff1983} approach in which the current $I(V)$ measured in STM is proportional to the integrated local density of states (LDOS) of the surface using the bSKAN code\cite{Hofer2003}.

\begin{figure}
\includegraphics[scale=1]{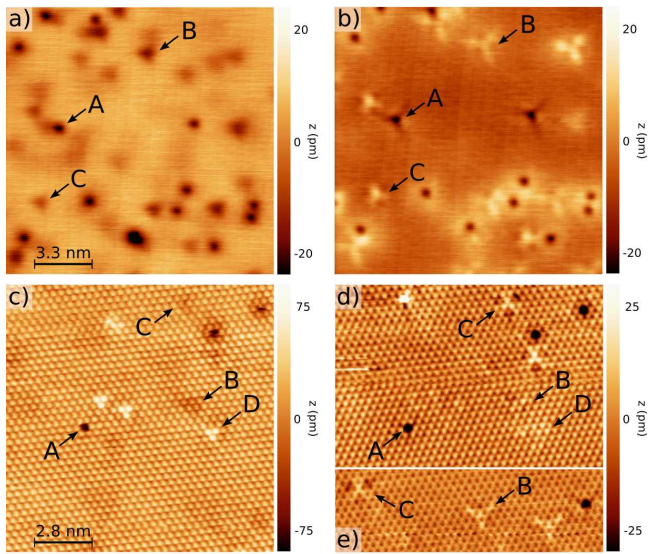}
\caption{\label{fig1}(color online). Filled and empty-state STM images of the same area of stoichiometric $1T$-TiSe$_2$ grown at 575\celsius (a,b) and of a Ti self-doped $1T$-TiSe$_2$ grown at 650\celsius (c,d). STM image (e) is from the same surface as (c,d) with another tip termination. Bias voltages : (a,c) -1V and (b,d,e) +1V. $I_{\text{t}}=0.2$ nA, $T=4.7$ K. Different defects are labeled A, B, C, D.}
\end{figure}
The STM images of the $1T$-TiSe$_2$ surface recorded at -1V (Fig. \ref{fig1}(a)) and +1V (Fig. \ref{fig1}(b)) reveal three distinct native defects characteristic of our stoichiometric crystal. Ti self-doped crystal synthesized at higher temperatures reveal a fourth kind of defect (Fig. \ref{fig1}(c)). The charge density wave is not resolved in these images because integrating electronic states within 1 eV make specific CDW contributions negligible\footnote{The CDW gap (smaller than 50 meV) opens just below the Fermi energy and the effects of backfolding of p-derived bands due to the new Brillouin zone are only observable close to the Fermi energy. The detailed discussion on the perturbation of the CDW state by the different defects is beyond the scope of this paper.}.

On stoichiometric $1T$-TiSe$_2$, all defects (A, B, C) can be observed as depletions in filled-state images (Fig. \ref{fig1}(a)) with some difference in shape and darkness \mbox{($\Delta z_{\text{A}}$ $>$ $\Delta z_{\text{B}}$ $>$ $\Delta z_{\text{C}}$)}. On the Ti self-doped samples, one additional kind of defect appears as a bright set of three atomic sites labeled D in Fig. \ref{fig1}(c). It may easily and directly be related to the excess of titanium in Ti$_{1+\text{x}}$Se$_2$. In fact, only these defects correspond to electron donor\footnote{In this study, the donor and acceptor terms generally used for (in)gap states induced by respectively n- and p-dopants in semiconductors have to be interpreted as valence/conduction impurity states.} defects and, as expected, the density of "bright" defects is clearly higher on the crystal grown at 900\celsius$~$ than in the 650\celsius$~$ one. 

If each protrusion/depletion is assigned to one atomic defect, a statistical estimation gives a density of 1-2$\%$ of native defects in the stoichiometric sample. Depending on the doping nature of these defects, this concentration can strongly affect the electronic properties. Solely on the basis of filled-state images (Fig. \ref{fig1}(a,c)), all of these intrinsic structural defects cannot clearly be distinguished except for defect D. The discrimination can only be done with the help of the empty-state images (Fig. \ref{fig1}(b,d,e)). Here we can clearly observe that defect A appears like a hole and defect C presents a bright central spot surrounded by 3 depletions. Probing defect B, which corresponds to a bright three fold star (Fig. \ref{fig1}(b)), is much more dependent on the tip sensitivity to its associated orbitals. Whereas its observation is difficult in Fig. \ref{fig1}(d), we can clearly observe its three fold star shape in Fig. \ref{fig1}(e) acquired on the same surface after a slight modification of the tip termination. The relative orientation between defects B and C is found to be identical to the one in Fig \ref{fig1}(b).
 
\begin{figure}
\includegraphics[scale=1]{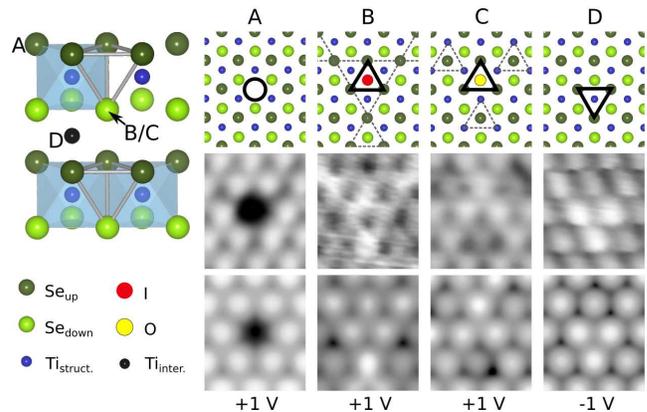}
\caption{\label{fig2}(color online). Atomically resolved STM images (central row) (1.23$\times$1.06 nm$^2$, $I_{\text{t}}$=200 pA, $T$=4.6 K), DFT simulated STM images (bottom row), structural inset\cite{Wilson1969} (top left) and schematic representation  (top row) for the four kinds of native defects in $1T$-TiSe$_2$. (A) Missing Se$_{\text{up}}$ atom (Se in top layer), (B) Substitution of Se$_{\text{down}}$ atom by an iodine atom,  (C) Substitution of Se$_{\text{down}}$ atom by an oxygen atom, (D) excess Ti intercalation.}
\end{figure}

Figure \ref{fig2} shows a zoom-in on the STM images with atomic resolution of Fig. \ref{fig1} (c,d,e), allowing to determine the registry of the defects with respect to the crystal lattice in order to compare them to simulated images and to precisely find out the origin (nature and conformation) of these four kinds of native defects.

On $1T$-TiSe$_2$, the topmost layer consists of Se atoms, whose $p$-orbitals contribute most to the tunneling current in filled-state images.\cite{Slough1988} Since we observe exactly the same atomic symmetry and positioning in the empty-state and in the filled-state STM images (Fig. \ref{fig1}(c,d)), we therefore conclude that the atoms observed on the images of Fig. \ref{fig2} can be uniquely attributed to Se atoms (Se$_\text{up}$). Depending on the doping nature (donor / acceptor) of the native defects, we have chosen to only present the most significant bias voltage for characterizing them in \mbox{Fig. \ref{fig2}}. Basically, filled- and empty-state STM images are used for highlighting, respectively, electron donors and acceptors\footnote{When not just due to a simple topographic effect, protrusions/depletions can correspond to donor/acceptor states\cite{Loth2008}.}. Corresponding simulated STM images (bottom row) and the schematic representation (top row) of the structural conformation (top view) are added for each defect of Fig. \ref{fig2}.

The atomically resolved STM image of defect A (Fig. \ref{fig2}(A) center) clearly shows that the depletion perfectly matches with the atomic position of a Se atom in the outermost layer (Se$_\text{up}$). By appearing as a depletion at all bias voltages between -1V and +1V, this defect obviously corresponds to a hole in topography and, in this way, can be associated with a Se vacancy in the outermost Se layer. The calculated STM image of a Se vacancy in this layer (Fig. \ref{fig2}(A) bottom) is in excellent agreement with the experiment.

Defects B and C are centered at the atomic position of a Se atom of the second Se layer (Se$_{\text{down}}$), just above the Van der Waals (VdW) gap (Fig. \ref{fig2}(B),(C)). Both exhibit depletions at -1V but have different redistribution of the electronic density at +1V. Indeed, in empty-state STM images defect B exhibits a 3-fold protrusion. Thus this kind of defect clearly has an electron acceptor behavior because it appears bright in empty-state images (see Fig. \ref{fig1}(b,e), \ref{fig2}(B)). Knowing that iodine is used for the preparation of the crystals and that it has been shown that, independently of the growth temperature, the $1T$-TiSe$_2$ crystals contain a residual iodine concentration of around \mbox{0.3 at. $\%$\cite{salvo1976}} we suppose that one Se atom of the second layer is substituted by an iodine atom. Furthermore, iodine is known to be more electronegative than selenium. In this way, due to the existence of a stronger electronegative species on the Se site in the second layer, the probability to inject electrons above this defect is slightly increased. In addition, as a consequence of this more electronegative element in the structure, the 3 neighboring Ti atoms also have a lower negative charge. This then affects the charge transfer to the other Se atoms since they should have less negative charge than the usual ones. The scanning tunneling microscope is more sensitive to the topmost layer and therefore it observes these modifications of the charge transfer via an increased probability of filling electrons on these Se atoms (appearing as protrusions in STM images at +1V). The simulation of an I-substitutional shows the same general behavior as the measured defect(see simulation of Fig. \ref{fig2}(B)). However, because of the limited size of the cluster used for calculation and the fact that the DFT method can not fully take into account the spatial extension of effects due to long range Coulomb interaction, the lateral size of the simulation is slightly smaller than the measured effect.

Defect C is also centered on top of a Se site of the second layer (Se$_{\text{down}}$) but displays a different pattern than defect B. In empty-state STM images (see Fig. \ref{fig1}(b,d,e), Fig. \ref{fig2}(C)), defect C is recognizable with a strong increase of the density of states localized on three neighboring Se atoms of the topmost layer and also 3 surrounding depletions. This defect can also be understood in terms of a substitution of a Se$_{\text{down}}$ atom by a more electronegative atom. Thus, oxygen is a good candidate for the reason that it has the same valence as Se and is an inevitable impurity of the atmosphere of the reaction system. In addition, it explains why defects B present a higher density than defects of type C on all of our $1T$-TiSe$_2$ samples. The simulation for O-substitutional (Fig. \ref{fig2}(C)) clearly confirms this supposition.

Note that we could not observe any substitution by iodine or oxygen in the topmost selenium layer (Se$_\text{up}$). This is probably due to the desorption of the least stable substituent shortly after cleaving at room temperature. This also explains why one only observes selenium vacancies in the topmost layer and not in the second selenium layer. 

As mentioned before, defects D can only be observed on Ti self-doped samples. In filled-states STM images (negative bias voltages below -0.2V), this defect appears as a bright protrusion (Fig. \ref{fig1}(c)), which corresponds in the atomically resolved image (Fig. \ref{fig2}(D)) to three neighboring Se atoms of the outermost layer appearing brighter. 
In empty-state STM images at positive bias voltages above 0.3V (Fig. \ref{fig1}(d)), the electronic perturbation induced by the defect is much lower and makes it nearly invisible. Therefore, defect D has a well-defined electron donor character, coinciding with the presence of an additional Ti atom which locally modifies the hybridization with neighboring Se atoms\cite{May2011a}. Total energy calculations show that additional Ti atoms should be placed in the VdW gap and in the alignment of Ti atoms from a top view (see schematic of Fig. \ref{fig2}(D)). The simulated STM image of an intercalated Ti atom in this conformation confirms our interpretation about defect D.

Finally, we point out that none of the four observed defects are due to a local excess of Se atoms in the structure. Furthermore, we could not associate any observed defect to Ti-vacancies.
\begin{figure}
\includegraphics[scale=1]{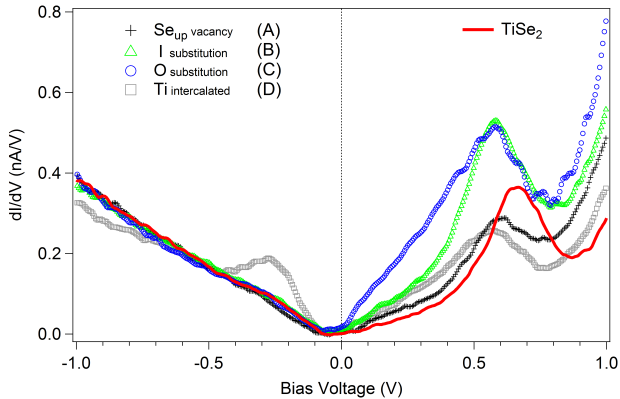}
\caption{\label{fig3}(color online). Experimental d$I$/d$V$ curves obtained on $1T$-TiSe$_2$ and on top of the different defects. The red curve is obtained far away from the observed defects.
(Averaging 10 spectra for each curve, $T$ = 4.6 K, V$_{\text{set}}$ =-1.1 V, $I_{\text{t}}$= 0.2 nA).}
\end{figure}

Figure \ref{fig3} presents d$I$/d$V$ curves obtained by STS measurements on the different types of defects and on the unperturbed surface, offering the possibility to visualize the local perturbation of the LDOS in the vicinity of a defect and to recognize the electronic properties of each of them as a function of energy. For the tip stabilization, $V_{\text{set}}$ was chosen to get a quasi-constant tip-surface distance for all STS curves. Indeed, the spectra have been recorded in the CDW phase (4.7 K) and one expects to see the opening of the CDW gap at the Fermi level (E$_\text{F}$). However, due to the presence of three conduction bands in $1T$-TiSe$_2$, the system opens a gap slightly below E$_\text{F}$\cite{Monney2009a}. The further discussion of CDW effects is beyond the scope of this paper.

Looking at these spectra, one can first observe that in the occupied states (negative bias voltage), all defects show almost the same behavior of the LDOS as defect-free regions of $1T$-TiSe$_2$, except for sites with intercalated Ti (defect D). We can conclude that defects A, B and C do not perturb strongly the Se $p$-bands below the Fermi level. In contrast, intercalated Ti atoms (defect D) present a well-defined peak in occupied states. This strong peak suggests the presence of a localized non-dispersing state that originates from the intercalated titanium $d$-orbital and explains why one observes the strong donor behaviour of this defect in filled-state STM images (Fig. \ref{fig1}(c), Fig. \ref{fig2}(D)).

Comparing the tunneling spectra of defect B (iodine substitution of Se$_{\text{down}}$) and the stoichiometric $1T$-TiSe$_2$ surface, it appears that an upward bending of the spectrum of the stoichiometric $1T$-TiSe$_2$ surface in the empty-state region can bring both spectra to a match. Such a bending may be interpreted in terms of a local band-bending of the structural titanium $d$-bands (explaining the strong acceptor behaviour of this defect), associated with a strong modification of the local electronic potential.

The spectrum acquired at the center of defect C (oxygen substitution of Se$_{\text{down}}$) presents an even more important increase of the density of states in the unoccupied states than defect B (attributed to iodine substitution). This is probably due to the fact that the presence of oxygen in the structure locally shifts the structural titanium $d$-bands towards the fermi-level because of its strong electronegativity. Thus, the density of states is already increased below E$_\text{F}$.

Finally, these five spectra allow the determination of the local spectroscopic signatures around the Fermi level of all associated native defects of $1T$-TiSe$_2$.

The structural and electronic properties of native defects in $1T$-TiSe$_2$ have been investigated. We have been able to precisely identify and characterize the origin and conformation of the four predominant atomic defects revealed by STM. They are intercalated Ti atoms, Se vacancies and Se substitution by residual iodine and oxygen atoms. This information will be helpful for further studies in particular for the investigation of the electronic effects of Ti self-doping on the CDW state, the temperature dependent conductance and for a better comprehension of the effect of extrinsic doping such as, for example, Cu intercalation leading to a superconducting state.

This project was supported by the Fonds National Suisse pour la Recherche Scientifique through Div. II . We would like to thank H. Beck, F. Vanini and C. Monney for motivating discussions. Skillful technical assistance was provided by G. Manfrini, F. Bourqui, B. Hediger and O. Raetzo.

\bibliography{library1}
\end{document}